\documentclass[conference]{IEEEtran}
\IEEEoverridecommandlockouts
\usepackage{xcolor}
\long\def\invis#1{}

\usepackage{amssymb}
\usepackage{mathtools}
\usepackage{graphicx}
\usepackage{float}
\usepackage{tikz}
\usepackage{listings}
\usepackage[colorlinks=true, linkcolor=red, citecolor=red, urlcolor=blue]{hyperref}

\usepackage{graphicx}
\usepackage{cite}
\usepackage{dsfont}
\usepackage{mathtools,etoolbox}
\usepackage[ruled, linesnumbered]{algorithm2e}
\usepackage[none]{hyphenat}
\usepackage[utf8]{inputenc}
\usepackage[english]{babel}
\usepackage[T1]{fontenc}
\usepackage{mathtools, nccmath}

\usepackage{amsthm}
\usepackage{xfrac}
\usepackage{nicefrac}
\usepackage{booktabs}
\usepackage[switch, pagewise]{lineno}
\usepackage{mathrsfs}
\DeclareMathAlphabet{\mathpzc}{OT1}{pzc}{m}{it}
\usepackage{multirow}
\usepackage{multicol}
\usepackage{bbm}
\usepackage{soul}
\usepackage{xfrac}
\usepackage{balance}
\usepackage{titlesec}
\usepackage[final]{pdfpages}
\usepackage{amsmath}
\usepackage{xcolor}

\usepackage{cite}
\usepackage{amsmath,amssymb,amsfonts}
\usepackage{algorithmic}
\usepackage{graphicx}
\usepackage{mathtools}
\usepackage{textcomp}
\usepackage{xcolor}
\usepackage{url}
\usepackage[affil-it]{authblk}
\def\BibTeX{{\rm B\kern-.05em{\sc i\kern-.025em b}\kern-.08em
    T\kern-.1667em\lower.7ex\hbox{E}\kern-.125emX}}

\usepackage{caption}

\begin{document}

\title{Performance Assessment of Feature Detection Methods for 2-D FS Sonar Imagery*\\
\thanks{This work was supported by USDA NIFA sustainable agriculture system program under award
number 20206801231805.}% <-this % stops a space
\thanks{}
}
\author{Hitesh Kyatham$^{1}$, Shahriar Negahdaripour$^{2}$, Michael Xu$^{1}$, Xiaomin Lin$^{1}$, \\
Miao Yu$^{1}$, Yiannis Aloimonos$^{1}$, 
}
\affil{$^{1}$Maryland Robotics Center, University of Maryland}
% \affil{$^{2}$University of Colorado Boulder}
\affil{$^{2}$ECE Department, University of Miami, Coral Gables, FL}
\maketitle
\begin{abstract}
Underwater robot perception is crucial in scientific subsea exploration and commercial operations. The key challenges include non-uniform lighting and poor visibility in turbid environments. High-frequency forward-look sonar cameras address these issues, by providing high-resolution imagery at maximum range of tens of meters, despite complexities posed by high degree of speckle noise, and lack of color and texture. In particular, robust feature detection is an essential initial step for automated object recognition, localization, navigation, and 3-D mapping. Various local feature detectors developed for RGB images are not well-suited for sonar data. To assess their performances, we evaluate a number of feature detectors using real sonar images from five different sonar devices. Performance metrics such as detection accuracy, false positives, and robustness to variations in target characteristics and sonar devices are applied to analyze the experimental results. The study would provide a deeper insight into the bottlenecks of feature detection for sonar data, and developing more effective methods

\end{abstract}

\section{sec:intro}
Nowadays underwater exploration and monitoring in fields such as marine archaeology and offshore oil operations heavily rely on the deployment of underwater robots. These robots require robust perception capabilities to be able to navigate and operate autonomously in challenging subsea environments. The adverse effects of nonuniform lighting and poor visibility in turbid waters, which limit the effectiveness of optical imaging systems. These are overcome by emerging 2-D high-frequency forward-look (FL) sonar cameras as essential tools for high-resolution imagery at ranges of up to about 50 meters, far beyond the capabilities of traditional optical cameras in even the clearest waters.

Some key challenges must be overcome in the automated processing of the data acquired by these imaging sonars: high levels of speckle noise, lack of color and texture information, and the inherent ambiguities as a ranging device.

\begin{figure}
    \centering
    \includegraphics[width=1\linewidth]{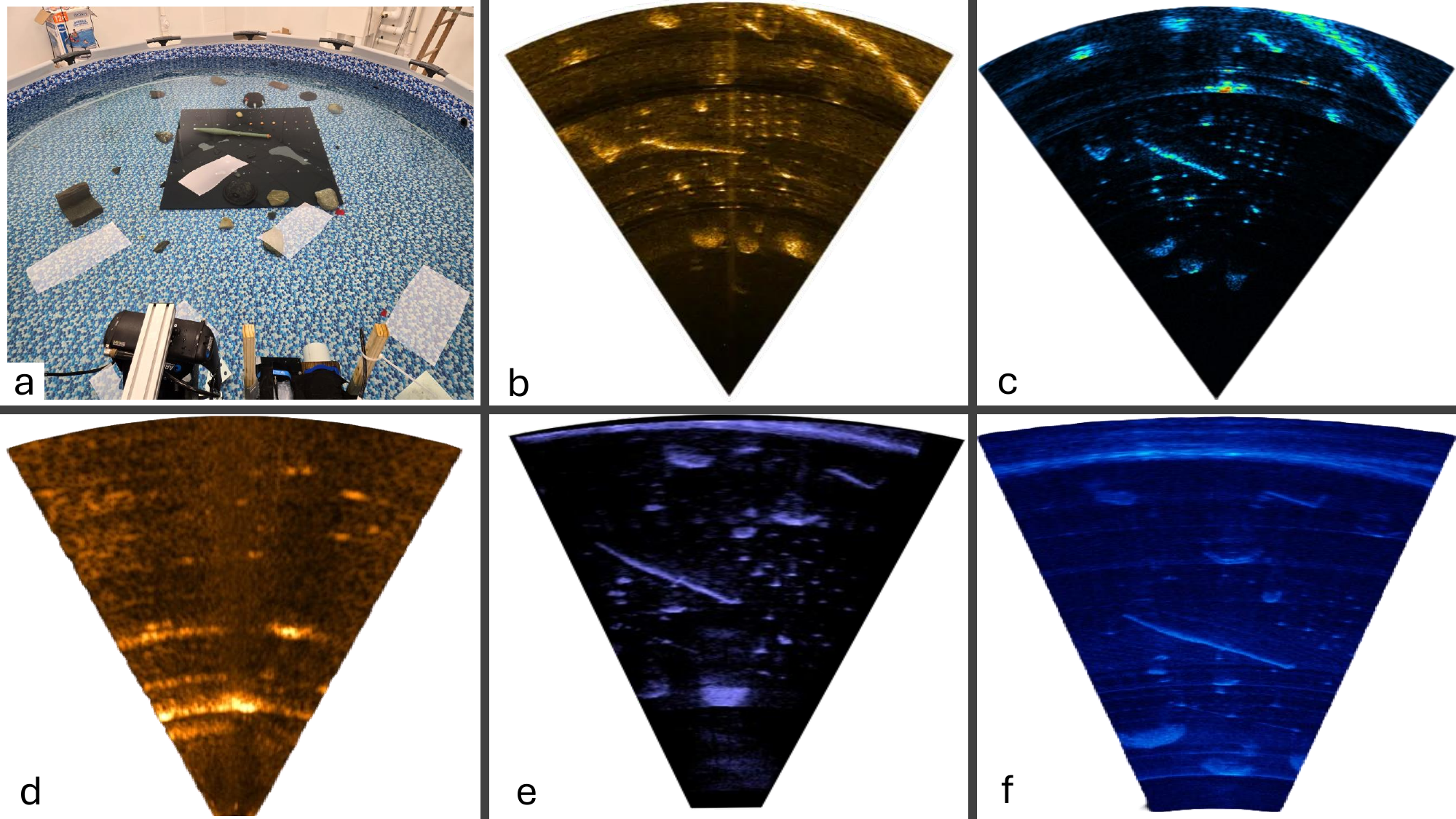}
    \caption{Overview of the setup and sample sonar images: (a) Experimental setup, (b) Oculus, (c) Gemini, (d) BlueView, (e) Didson, (f) Aris.}
    \label{fig:sonars}
    \vspace{-5mm}
\end{figure}

Automatic feature detection is a fundamental step in sonar image processing for various perceptual tasks, e.g., object recognition, precision landmark-based localization, navigation, re-acquisition, and 3-D reconstruction. However, there has been little progress in the development of effective local feature detectors that are tailored to the unique characteristics of the sonar data. Current approaches in image registration, such as global methods resistant to speckle noise \cite{aykin2012feature, johannsson2010imaging} or machine learning techniques that leverage example-based learning \cite{de2017end, balakrishnan2019voxelmorph}, have not fully addressed the need for robust feature detection in sonar imagery. 

Some key obstacles arise from previous evaluations of feature detection in sonar images have been limited in scope, often relying on synthesized data rather than real-world sonar imagery. This investigation seeks to provide a comprehensive evaluation of existing local feature detection methods, including well-known techniques such as SIFT, SURF, FAST, ORB, BRISK, SU-BRISK, F-SIFT, and KAZE \cite{SIFT_Tutorial,bay2006surf, alcantarilla2011fast, lowe2004sift, alcantarilla2012kaze,rublee2011orb, leutenegger2011brisk, ozuysal2009fast,zhao2012flip, tareen2018comparative} on real data. Moreover, the performance assessment is extended further by analyzing the data acquired with different sonar devices for the same target types under similar environmental conditions. Among key factors are the number of detected features identified by different detectors in each sonar image type, and repeatability in terms of detecting the same features in the image. In particular, we provide answers to two key questions: does a particular method stands out, and does a particular sonar yields more features across most, if not detectors (stay tuned).

By systematically analyzing these findings, we aim to facilitate the development of more effective feature detection methods for sonar data, ultimately enhancing the underwater robot perception for a wide range of capabilities: automated positioning and navigation, docking, mapping, target detection, and recognition.

The remainder of this paper is organized as follows. In Section~\ref{sec:related_work}, we discuss recent advances and related work. We discuss how we performed our experiments in Section~\ref{sec:experiments}. We present qualitative and quantitative analysis in Section~\ref{sec:results}. We conclude our with final comments in Section~\ref{sec:conclusion}.
\section{Related Work}
\label{sec:related_work}

Feature detection is a crucial method in computer vision, widely used to identify distinct parts in an image, to aid in recognizing various patterns, and to extract relevant information that facilitates data interpretation. In our study, we employed eight well-known feature detectors: SIFT, SURF, FAST, ORB, BRISK, SU-BRISK, F-SIFT, and KAZE \cite{SIFT_Tutorial, bay2006surf, alcantarilla2011fast, lowe2004sift, alcantarilla2012kaze, rublee2011orb, leutenegger2011brisk, ozuysal2009fast, zhao2012flip, tareen2018comparative}. These detectors were chosen for their higher performance and availability in open-source packages such as OpenCV.

Autonomous underwater vehicles (AUVs) depend on robust feature detection algorithms for accurate navigation and mapping of underwater environments. One such application is AcTag, a new fiducial marker system designed for underwater environments that combines visual and acoustic markers to enhance the localization and mapping capabilities of robotic systems \cite{norman2023actag}. However, there are limitations specific to AcTags in the context of feature detection. For example, while AcTags provide multiple unique landmarks, the number of distinguishable features may be limited compared to other fiducial markers. Additionally, the effectiveness of feature detection is highly dependent on the resolution of the sonar system used; lower resolution sonar may not capture the fine details of the AcTag, leading to challenges in accurately detecting and interpreting the features. These limitations highlight the need for careful consideration of the operational environment and the capabilities of the sonar system to ensure effective detection and localization.
\begin{figure}
    \centering
    \includegraphics[width=1\linewidth]{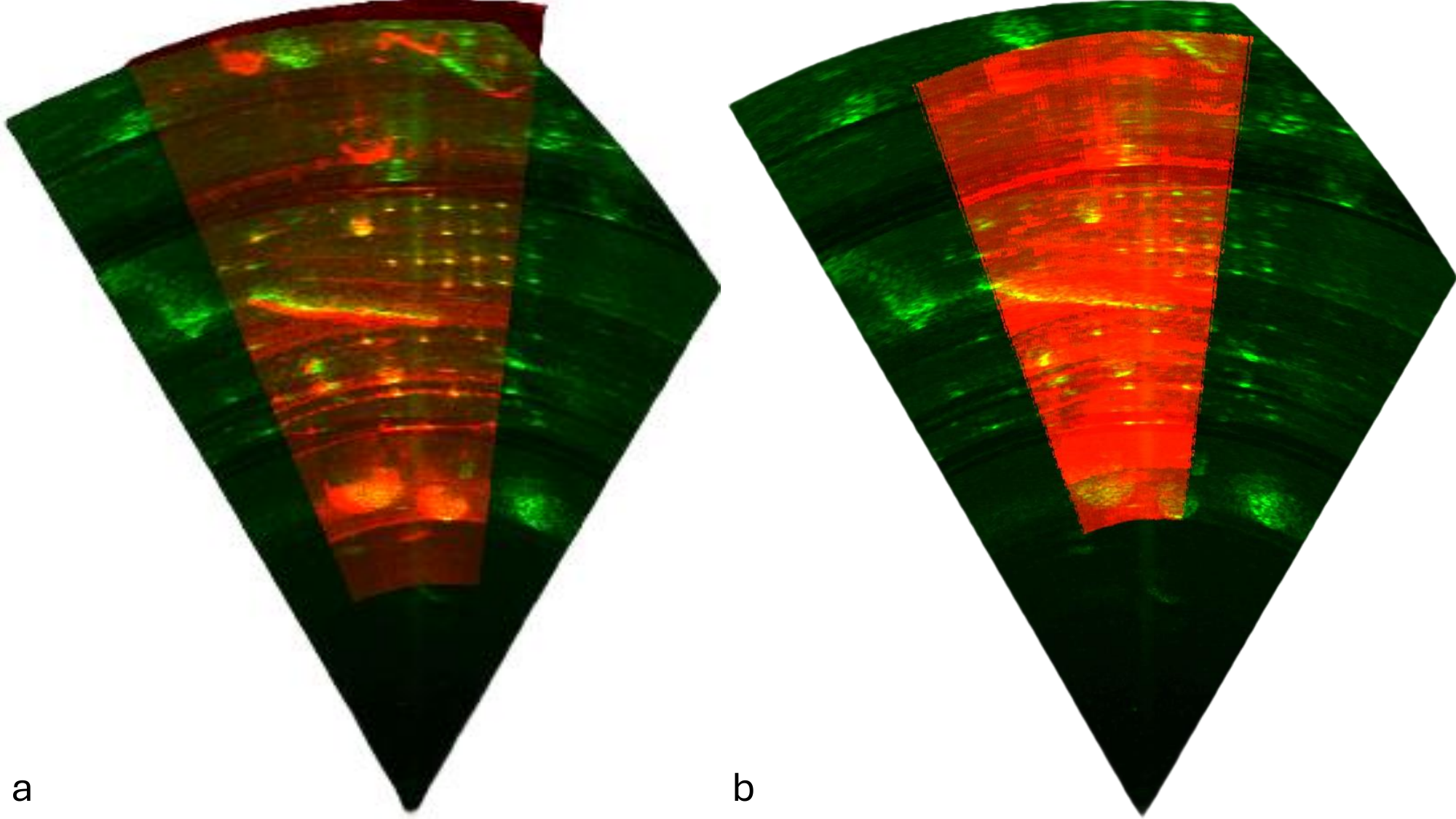}
    \caption{Lens distortion in lens-based Didson and Aris sonar has to be rectified to satisfy ideal sonar projection model (a) distorted Aris image (red channel) does not aligned correctly with overlapping Oculus image (green channel). (b) two image align correctly after distortion correction with a mild cubic model.}
    \label{fig:DistortedVsUnDistorted}
\end{figure}

Previous work used a synthetic dataset of 600 sonar images \cite{tueller2018comparison}, each featuring a single mine line object superimposed on three different backgrounds. The performance of feature detection algorithms are assessed using the Support Vector Machine (SVM) classifier. The effectiveness of these detectors was measured using receiver operating characteristic (ROC) curves, by plotting the true positive rate against the false positive rate. The results indicated that while some detectors, such as SURF, performed well in specific scenarios (e.g. sand backgrounds), others, such as FAST, exhibited consistent performance across all datasets. This study underscored the challenges of feature detection in underwater environments, suggesting that further research should involve real sonar images for a more comprehensive analysis.

The reliance on computer-generated data in the previous work limited its scope, and falls short of meeting the objectives of our investigation. Our research provides a more comprehensive analysis by applying the feature detection methods to real sonar images from 5 different FL sonar devices: Aris Explorer 3000, BlueView M900, dual-frequency DIDSON, Gemini 1200ik, and Oculus M1200d sonars.

\section{Experiment and Analysis}
\label{sec:experiments}
% \XL{You probably need a paragraph here to go from related work to experiment. Why is it important and why do we need to do our experiments?}
Building on the limitations and findings from previous studies, this research aims to bridge these gaps by utilizing real sonar images from five distinct FL sonar devices: Aris Explorer 3000, BlueView M900, dual-frequency DIDSON, Gemini 1200ik, and Oculus M1200d. This methodology provides a more robust evaluation of feature detection algorithms in authentic underwater settings. By employing real-world data, this study seeks to enhance the understanding of these algorithms' performance and their practical applications in underwater exploration and navigation. The subsequent section details the experimental setup and the dataset employed for our analysis.
\subsection{Setup and Dataset}

\begin{figure*}[ht!]
    \centering
    \includegraphics[width=1\textwidth]{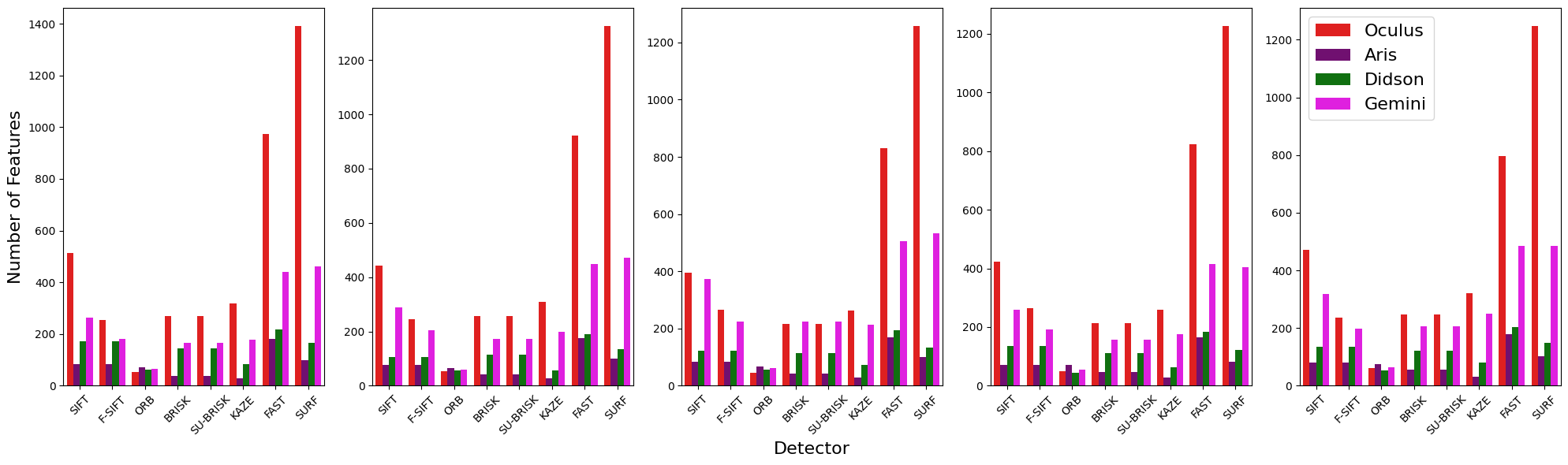}
    \caption{First data set comparative analysis of feature detection performance among eight detectors (SIFT, SURF, FAST, ORB, BRISK, SU-BRISK, F-SIFT, and KAZE) using four sonars (Oculus, Aris, Didson, and Gemini). Bar graph illustrates the number of features detected by each sonar-detector combination for a fixed sonar position. Each of the five subplots, represents a distinct sonar position.}
    \label{fig:Feature_count_1}
\end{figure*}

There are hundreds if not thousands of freely available RGB datasets(e.g., COCO \cite{lin2014microsoft} and ImageNet \cite{deng2009imagenet}), for the analysis and assessment of various machine vision tasks. However, acquiring real sonar data is challenging due to the high cost of sonar equipment and their deployment. In addition to the three types of sonar (Didson \cite{Soundmetrics}, BlueView \cite{Teledynemarine}, and Oculus \cite{Blueprintsubsea}) we possess, we reached out and described our project to two sonar companies (Soundmetrics \cite{Soundmetrics} and Tritech \cite{Tritech}) for Aris and Gemini loaner units. Both companies accommodated our requests without hesitation. 
% Additionally, Soundmetrics offered to operate the devices for us, and the company's CEO and technical support member attended and operated the Aris for data collection. 

As depicted in Fig.~\ref{fig:sonars}\textcolor{red}{a}, the experimental setup is an indoor water tank, approximately 1 meter deep and 4 meters in diameter. The targets include a semitransparent planar board with various small features of different shapes and sizes: spherical and relatively flat pieces of random shapes, glued to the board at various regular spacings. In addition, the setup depicted in the figure includes a torpedo-shaped mine-shaped object, two stacked mine-shaped targets, and rocks of various sizes. We positioned the sonars side by side to ensure roughly similar conditions in capturing the data.

In our first data set, the position of the feature board is varied (in each of 5 poses) while keeping the sonars and other targets stationary. Moreover, speckle noise is significantly reduced by averaging over 9 frames of ``still video.’’ \footnote{In this part only, BlueView sonar was not available and not utilized.}.

In the second dataset, the feature board and other targets (minus the torpedo-shaped object) are fixed in place, while each sonar moves along the boundary of the pool (one by one) to capture a video. Thus, the complete data set is made up of video images at five different positions (first part) and video data for the second part. For the confined scope of this study, we identified and selected five positions (from each video) with the most similar poses among all five sonars.

To reduce the impact of speckle noise at the expense of minimal motion blur, we performed temporal averaging over 5 frames centered at each pose.

The primary difference between the two datasets is the frame averaging strategy: in the second dataset, the five-frame moving average reduces the speckle noise with minimal motion blur. In contrast, there is no motion blur in the first set, and only nine-frame averaging results in more immunity to speckle noise, yet not significantly different from the second set.

% \XL{Likewise, one sentence here to go to the next assessment nad method}
\begin{figure*}
    \vspace{3mm}
    \includegraphics[width=\textwidth]{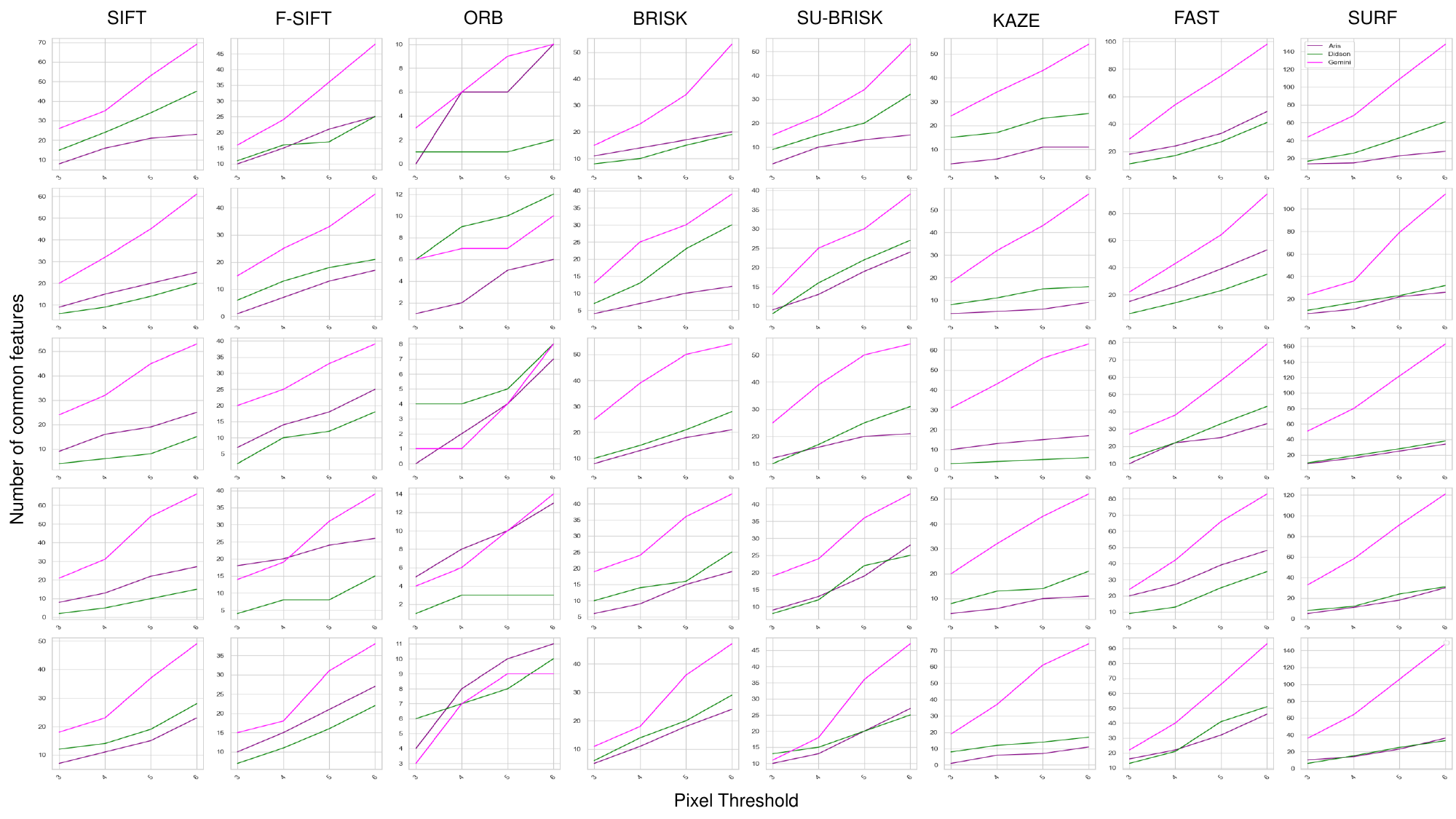}
    \centering
    \caption{First dataset - assessing effectiveness of different detector in identifying common features in images of different sonar types. The y-axis in each graph represents the number of common features in sonar pairs, showing comparisons of Oculus vs. Aris (purple line), Oculus vs. Didson (green line), and Oculus vs. Gemini (magenta line). The x-axis represents the distance threshold [pix]. Each of the 8 columns in the array corresponds to a different position.}
    \vspace{-2mm}
    \label{fig:Common_Part_1}
\end{figure*}

\subsection{Assessment Methodology}
To perform a comprehensive feature detection performance analysis, we need to establish a ground truth for comparison. Due to various factors, a sufficiently accurate ground truth cannot be ensured. Instead, we processed the sonar images through all eight feature detectors (SIFT, F-SIFT, ORB, BRISK, SU-BRISK, KAZE, FAST, and SURF) and observed that the Oculus sonar images had the most number of detected features by far \footnote{This is partly due to narrow horizontal field of view (roughly 28$^\circ$) for Didson and Aris, and roughly 45$ ^\circ $ for BlueView. In comparison, Oculus and Gemini have a 60-degree FoV}. Loosely speaking, the Oculus features are selected as ground truth (solely due to a higher number of features). In place of detection error, namely, the distance of each detected feature from the ground truth, we measure (for each sonar) the number $ N_d $ of features at distance $d=3$ to 6 pixels from an Oculus sonar feature.

Moreover, we apply an affine transformation to align each other sonar image with the Oculus view\footnote{Sonar image transformation under rigid body motion depends on the scene geometry, but can often be approximated by a fixed affine model with reasonable accuracy.}. Finally, we have applied a non-minimal supression with a 10-pixel threshold to discard multiple features in the neighbourhood of the most significant feature.

\subsection{Pre-Processing}
The lens-based Didson and Aris sonars suffer lens distortion, primarily in the azimuth direction; see Fig.~\ref{fig:DistortedVsUnDistorted}. In general, we also have to correct for any deviations from the assumed sound speed in water. Both require an internal calibration process: to determine the mapping from each beam to the corresponding azimuth angle; to adjust the range measurements by a constant scaling. The former involves a mild cubic transformation, while no range adjustment has been necessary.

As stated, we assess the feature counts of each image based on the similarity with Oculus features, using Euclidean distance thresholds of 3, 4, 5, and 6 pixels.This approach allows us to quantify the performance of each feature detector across different sonar images without ground truth, providing a fair basis for comparison.
\section{Results}
\label{sec:results}  
\begin{figure}
    \centering
    \includegraphics[width=1\linewidth]{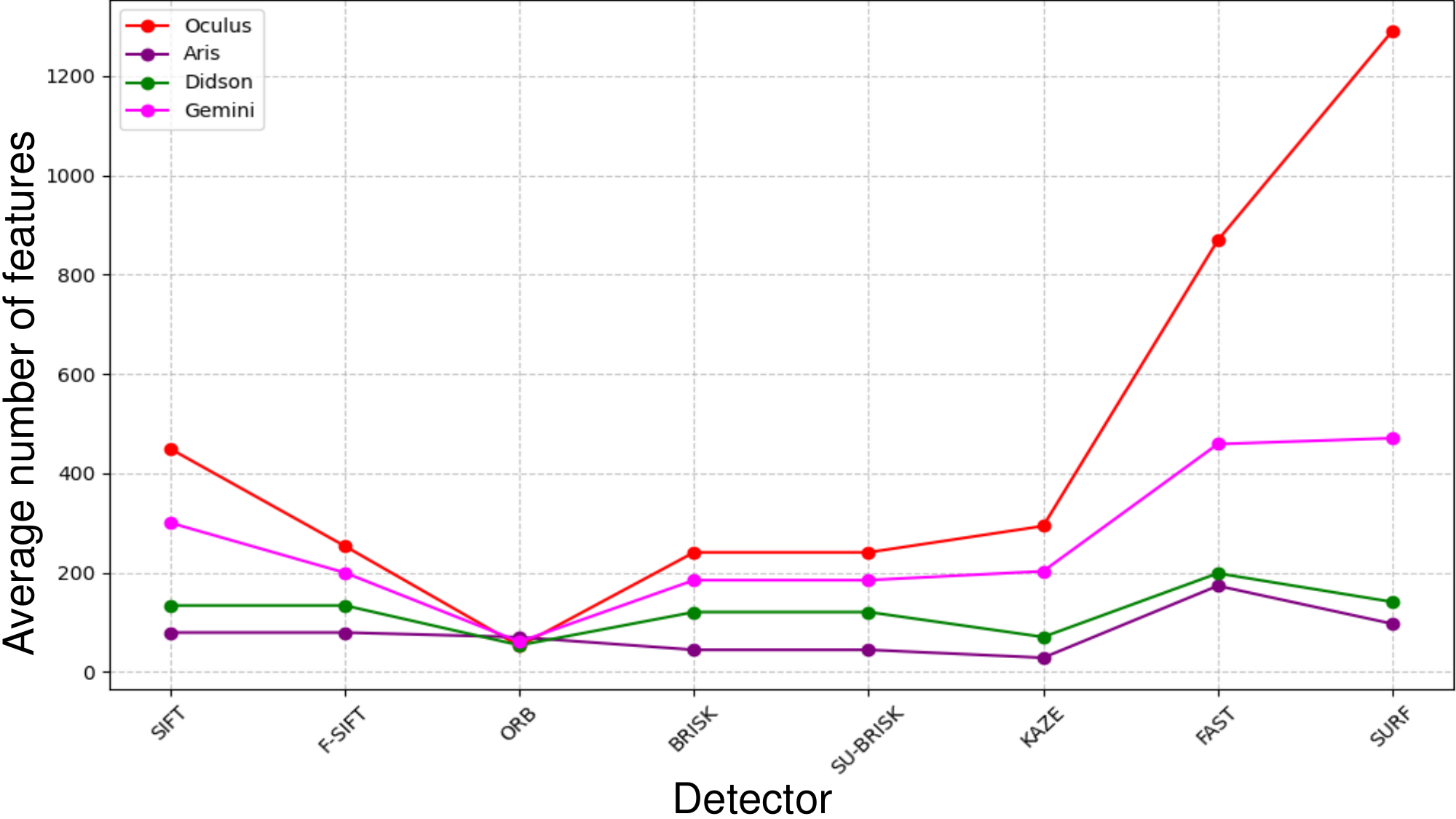}
    \caption{First dataset - average number of features detected by various sonar systems (Oculus, Aris, Didson and Gemini) for different feature detection algorithms.}
    \label{fig:Average_part1}
\end{figure}
\begin{figure*}
    \centering
    \includegraphics[width=1\textwidth]{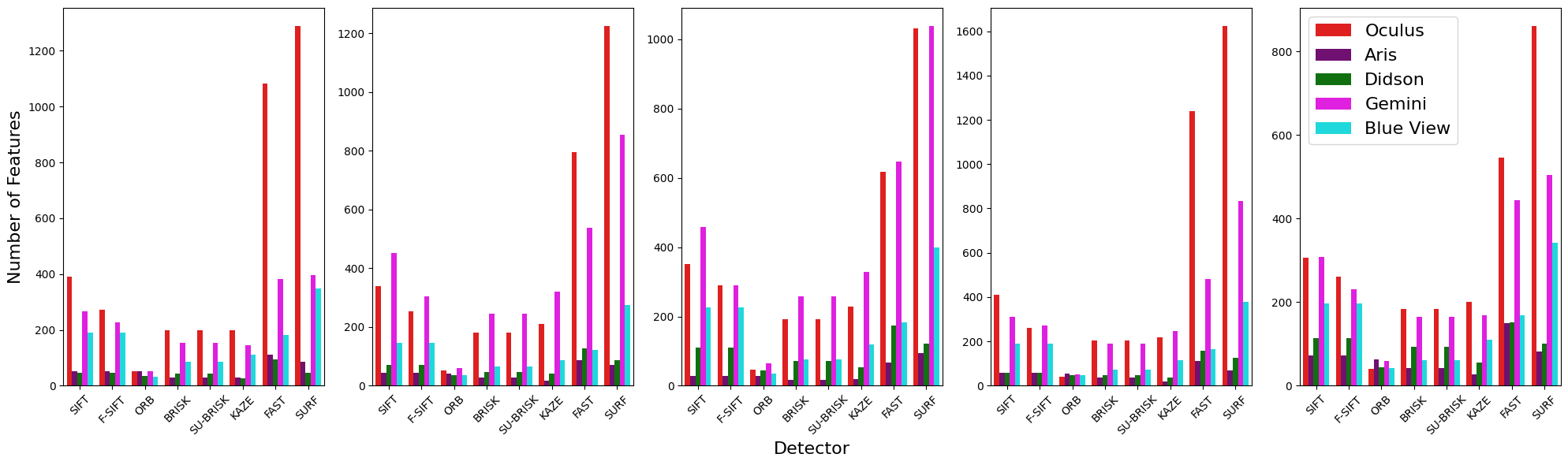}
    \caption{Second data set comparative analysis of feature detection performance among eight detectors (SIFT, SURF, FAST, ORB, BRISK, SU-BRISK, F-SIFT, and KAZE) using four sonars (Oculus, Aris, Didson, and Gemini). Bar graph illustrates the number of features detected by each combination of sonar detectors for varying sonar position. Each of the five subplots, represents a distinct sonar position.}
    \label{fig:Feature_Count_2}
\end{figure*}
This section presents the results of our experimental analysis, concentrating on the performance of various feature detection algorithms across different sonar systems and datasets. The evaluation entails a thorough comparison of the number of detected features and the consistency of feature detection across various methods and conditions. The results are systematically organized into two main parts, corresponding to the two datasets described above.
\subsection{First Dataset}

Fig.~\ref{fig:Feature_count_1} presents a bar graph to compare the number of features detected by four sonar systems- Oculus, Aris, Didson, and Gemini- across various feature detection methods. Each subplot corresponds to one of the five distinct poses in the dataset. The Oculus sonar consistently outperforms the other systems in the number of detected features across nearly all methods and positions. Here, the highest number of features detected is 1,392, achieved by Oculus sonar with the SURF feature detector, while the lowest is 27, recorded by the Aris sonar with the KAZE feature detector. There is a considerable variation in the number of detected features across different feature detection methods for each sonar system, with the KAZE feature detector notably yielding a low feature count for the Aris data in comparison to other Sonars.

Fig.~\ref{fig:Common_Part_1} illustrates a matrix of line graphs for the variation in the number of so-called "common features" by each feature detector based on a distance threshold of \(d=3\)[pix] to 6 pixels. Each row in the matrix represents a specific pose (over all 8 detectors), while each column corresponds to a particular feature detector (for all 5 poses). For example,  a maximum number of 165 common features are detected in a Gemini image (pose-three) with the SURF feature detector (for a distance threshold of $d=6$). In contrast, we obtain no common features in an Aris sonar image for the ORB feature detector (with $d=3$ in the first and third poses).

Fig.~\ref{fig:Average_part1} displays the average number of detected features for each of three sonars across all five positions and eight detectors. The average number of detected features ranges from 52 to 1,291. The Oculus system provides both the minimum and maximum average feature counts, corresponding to the ORB and SURF feature detectors, respectively.

\subsection{Second Dataset}
The bar graphs presented in Fig.~\ref{fig:Feature_Count_2} illustrate the number of detected features ($y$-axis) by various feature detectors ($x$-axis) at five selected positions along the sonar video tracks.

Again, the Oculus sonar image consistently yields the highest number of detected features, particularly with the FAST and SURF detectors. Feature counts in these cases often exceed 1000 and, reach up to 1400. The SIFT detector also demonstrates notable performance with Oculus sonar, detecting approximately 600 features at all positions. The performance in both Aris and Gemini sonar images is moderate, although there is still a higher count for the Gemini data. For example, Gemini sonar yields between 300 and 400 features when using the SIFT and SURF detectors, whereas Aris detects approximately 200 to 250 features with the same detectors. In contrast, the Didson sonar consistently records the fewest feature detections across all detectors and positions, with counts typically below 100 features.

\begin{figure*}
    \vspace{3mm}
    \includegraphics[width=\textwidth]{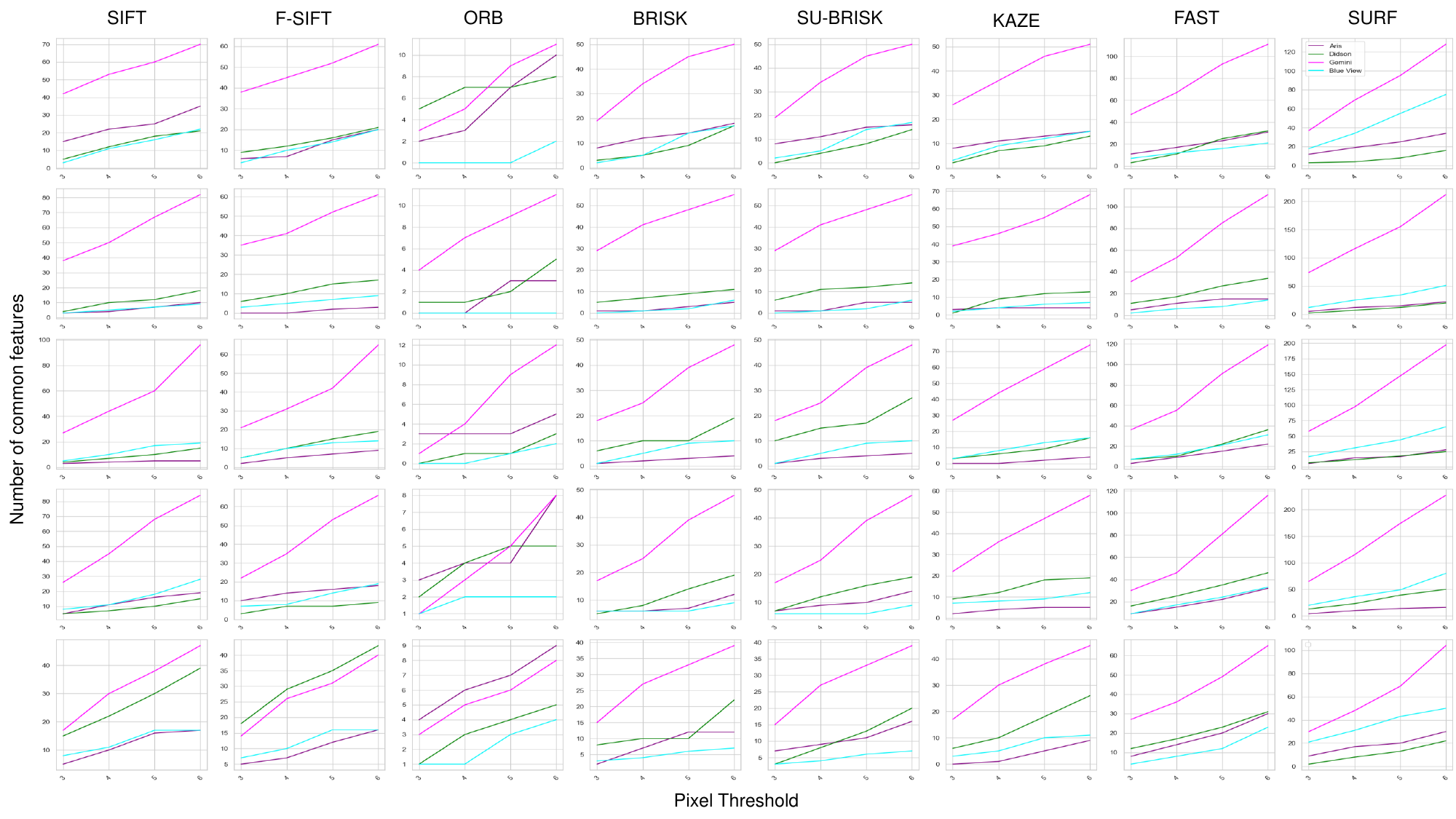}
    \centering
    \caption{Second dataset - assessing effectiveness of different detector in identifying common features in images of different sonar types. The y-axis in each graph represents the number of common features in sonar pairs, showing comparisons of Oculus vs. Aris (purple line), Oculus vs. Didson (green line), Oculus vs. Gemini (magenta line) and Oculus vs. BlueView (Cyan line). The x-axis represents the distance threshold [pix]. Each of the 8 columns in the array corresponds to a different position.}
    \vspace{-2mm}
    \label{fig:Common_Part_2}
\end{figure*}

Fig.~\ref{fig:Common_Part_2} presents a matrix of line graphs for variations in the number of common features for each detector as a function of distance, ranging from $d=3$ to $d=6$ [pix]. Each row in the matrix corresponds to a specific position, while each column represents a specific detector. The Gemini sonar, records the maximum number of common features, approximately 240, when utilizing the SURF detector. Conversely, we obtain from zero to less than handful of common features in the BlueView sonar images (over all positions), when employing the ORB feature detector.

\begin{figure}
    \centering
    \includegraphics[width=1\linewidth]{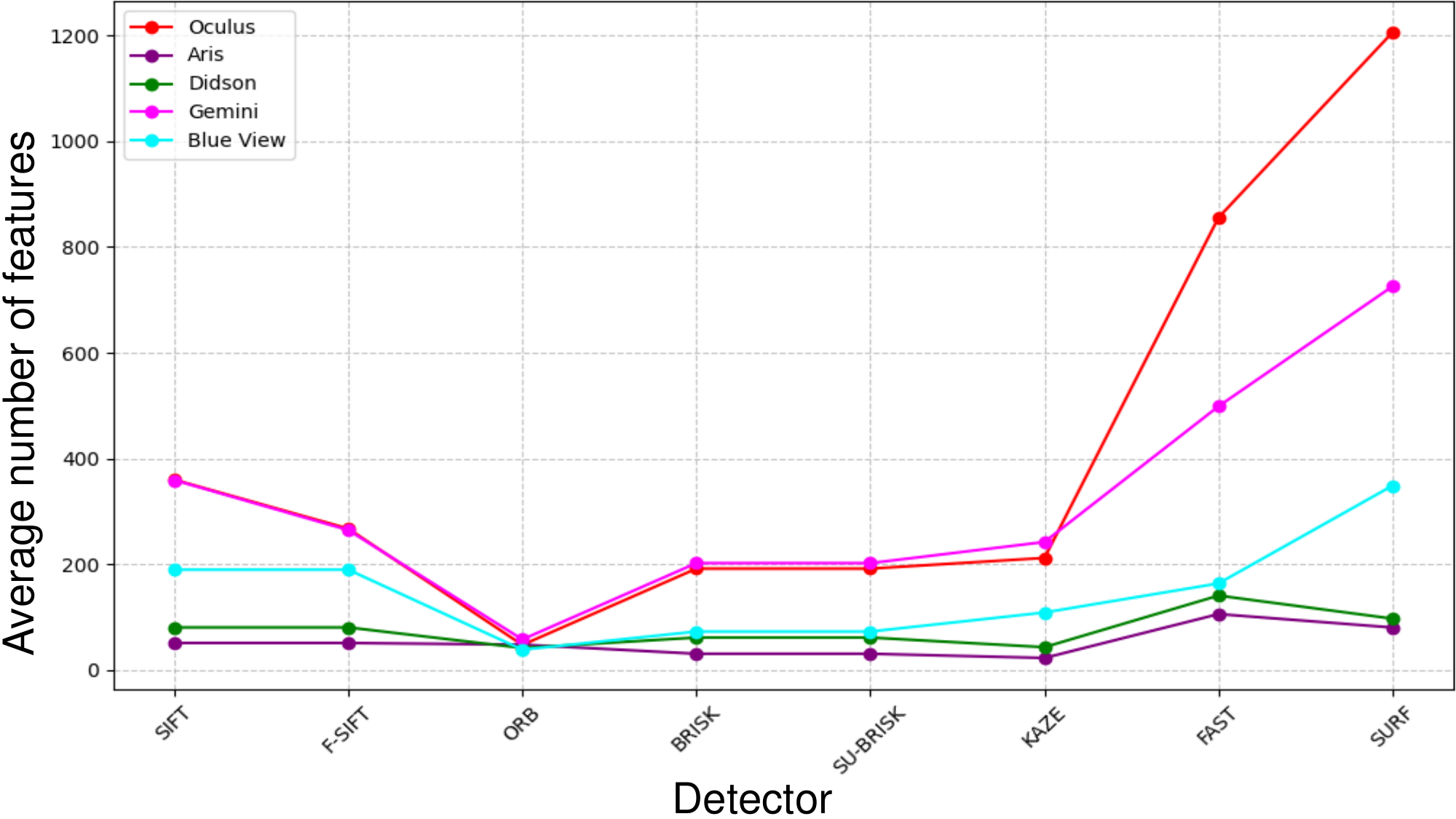}
    \caption{Second dataset - average number of features detected by various sonar systems (Oculus, Aris, Didson and Gemini) for different feature detection algorithms.}
    \label{fig:Average_part2}
\end{figure}

Fig.~\ref{fig:Average_part2} illustrates the average number of detected features by various sonar systems across all detectors. The performance of feature detectors varies significantly, from 22 to 1206, with the type of sonar and the detector used. The ORB detector consistently records the minimum number of features, whereas the SURF detector consistently yields the maximum average number of features across all sonar systems.

\section{Conclusions}
\label{sec:conclusion}
A summary of conclusions can be made in terms of various criteria, detector types, and sonar systems.

\subsection{Feature Count Analysis} 

The feature count highlights the most significant differences in the performance of feature detectors across the five types of sonar (Oculus, Aris, Didson, Gemini, and BlueView). Among, key findings, the Oculus images consistently produce a higher number of features across all detectors, with respect to the sonar type imagery. (mainly due to a higher scene coverage with 60$^\circ$ in horizontal FoV, also equaled by Gemini sonar). For this reason, the Gemini images consistently yield a higher number of common features across all detectors and thresholds.

BlueView Sonar with the next highest horizontal FoV (roughly 45$^\circ$) performs moderately, albeit for significantly smaller feature counts.

Aris and Didson, with their narrow horizontal FoV, yield the fewest number of features.

\subsection{Detector Performance}

The performance of individual feature detectors also varies, SURF consistently detects a higher number of features across all sonar types. However, one observation not explicitly substantiated in our current results is that its performance is less effective with small features. Although this is to be verified through a more focused investigation, it may suggest that SURF is more effective for images with larger-scale features and less for finer details.

FAST detector also consistently yields a higher number of features across all types of sonar.

SIFT and F-SIFT produce a moderate number of features (still less than FAST), but with variable performance.

ORB, BRISK, and SU-BRISK detect the fewest features overall, with KAZE performing slightly better (still far below the top-performing detectors).

\subsection{Common Feature Count Analysis} 

Ideally, a key criteria is the accuracy and effectiveness of a detector in identifying and localizing the same features across different sonar types. In the absence of ground-truth feature positions, we have employed the common feature count. Moreover, varying the distance thresholds of 3, 4, 5 and 6 pixels discounts the errors in aligning images from any sonar image with the Oculus (selected as reference due to a much higher feature count). Not surprising, more common  features are found when increasing the distance threshold. Generally, the results primarily support the earlier conclusion. For example, Gemini (with the second highest number of detected features) gives a higher number of common features, (likely due to similar azimuth angle range to the Oculus sonar).

Imperfect lens distortion correction in dual-frequency DIDSON and Aris Explorer 3000 sonar can introduce some feature localization error, thus partially contributing to a lesser number of common features.

\subsection{Final Comments}
We have provided only one sample image set in Fig.~\ref{fig:sonars}, for different sonar types, due to space limitation. Moreover, the complexity of establishing ground truth have prohibited the use of typical performance criteria applied to RGB images \cite{tareen2018comparative}. Still, the conclusions from this study are expected to provide a useful guideline in selecting feature detectors based on sonar type. In a future report, we plan to include details on the number, size, and reflectance properties of various scene features.

Further research could involve expanding the dataset to include a more diverse range of underwater environments and objects. In addition, collaborating with marine scientists and oceanographers to integrate biological and ecological factors into the analysis would incorporate application-specific and relevance factors, among others. In addition, developing adaptive algorithms that can adjust to underwater conditions, such as changes in water acoustic properties, would significantly advance the technology.

\section*{Acknowledgment}

At Soundmetrics, we thank Kimberly Bartinaux for earlier Aris loaner arrangments;  Jeanne Dorsey and Zach Duncan for attending  our data collection session and deploying the Aris sonar; Bill Hanot for technical support in Aris video image extraction. At Subsea Technologies: we thank Josh Dugan and Kellie Garden of various Gemini loaner arrangements, including the shipping.The first author would like to thank Divya Nallawar for assistance with data preparation and graphs. The travel expenses of the first author has been covered by a grant from USDA NIFA sustainable agriculture system program under award number 20206801231805.

%%%%%%%%%%%%%%%%%%%%%%%%%%%%%%%%%%%%%%%%%%%%%%%%%%%%%%%%%%%%%%%%%%%%%%%%%%%%%%%%
\bibliographystyle{IEEEtran}
\bibliography{references}

\begin{thebibliography}{10}
\providecommand{\url}[1]{#1}
\csname url@samestyle\endcsname
\providecommand{\newblock}{\relax}
\providecommand{\bibinfo}[2]{#2}
\providecommand{\BIBentrySTDinterwordspacing}{\spaceskip=0pt\relax}
\providecommand{\BIBentryALTinterwordstretchfactor}{4}
\providecommand{\BIBentryALTinterwordspacing}{\spaceskip=\fontdimen2\font plus
\BIBentryALTinterwordstretchfactor\fontdimen3\font minus \fontdimen4\font\relax}
\providecommand{\BIBforeignlanguage}[2]{{%
\expandafter\ifx\csname l@#1\endcsname\relax
\typeout{** WARNING: IEEEtran.bst: No hyphenation pattern has been}%
\typeout{** loaded for the language `#1'. Using the pattern for}%
\typeout{** the default language instead.}%
\else
\language=\csname l@#1\endcsname
\fi
#2}}
\providecommand{\BIBdecl}{\relax}
\BIBdecl

\bibitem{aykin2012feature}
M.~Aykin and S.~Negahdaripour, ``On feature extraction and region matching for forward scan sonar imaging,'' in \emph{2012 Oceans}.\hskip 1em plus 0.5em minus 0.4em\relax IEEE, 2012, pp. 1--9.

\bibitem{johannsson2010imaging}
H.~Johannsson, M.~Kaess, B.~Englot, F.~Hover, and J.~Leonard, ``Imaging sonar-aided navigation for autonomous underwater harbor surveillance,'' in \emph{2010 IEEE/RSJ International Conference on Intelligent Robots and Systems}.\hskip 1em plus 0.5em minus 0.4em\relax IEEE, 2010, pp. 4396--4403.

\bibitem{de2017end}
B.~D. De~Vos, F.~F. Berendsen, M.~A. Viergever, M.~Staring, and I.~I{\v{s}}gum, ``End-to-end unsupervised deformable image registration with a convolutional neural network,'' in \emph{Deep Learning in Medical Image Analysis and Multimodal Learning for Clinical Decision Support: Third International Workshop, DLMIA 2017, and 7th International Workshop, ML-CDS 2017, Held in Conjunction with MICCAI 2017, Qu{\'e}bec City, QC, Canada, September 14, Proceedings 3}.\hskip 1em plus 0.5em minus 0.4em\relax Springer, 2017, pp. 204--212.

\bibitem{balakrishnan2019voxelmorph}
G.~Balakrishnan, A.~Zhao, M.~R. Sabuncu, J.~Guttag, and A.~V. Dalca, ``Voxelmorph: a learning framework for deformable medical image registration,'' \emph{IEEE transactions on medical imaging}, vol.~38, no.~8, pp. 1788--1800, 2019.

\bibitem{SIFT_Tutorial}
\BIBentryALTinterwordspacing
opencv, \emph{opencv - Tutorial SIFT}, opencv. [Online]. Available: \url{https://docs.opencv.org/4.x/da/df5/tutorial_py_sift_intro.html}
\BIBentrySTDinterwordspacing

\bibitem{bay2006surf}
H.~Bay, T.~Tuytelaars, and L.~Van~Gool, ``Surf: Speeded up robust features,'' in \emph{Computer Vision--ECCV 2006: 9th European Conference on Computer Vision, Graz, Austria, May 7-13, 2006. Proceedings, Part I 9}.\hskip 1em plus 0.5em minus 0.4em\relax Springer, 2006, pp. 404--417.

\bibitem{alcantarilla2011fast}
P.~F. Alcantarilla and T.~Solutions, ``Fast explicit diffusion for accelerated features in nonlinear scale spaces,'' \emph{IEEE Trans. Patt. Anal. Mach. Intell}, vol.~34, no.~7, pp. 1281--1298, 2011.

\bibitem{lowe2004sift}
G.~Lowe, ``Sift-the scale invariant feature transform,'' \emph{Int. J}, vol.~2, no. 91-110, p.~2, 2004.

\bibitem{alcantarilla2012kaze}
P.~F. Alcantarilla, A.~Bartoli, and A.~J. Davison, ``Kaze features,'' in \emph{Computer Vision--ECCV 2012: 12th European Conference on Computer Vision, Florence, Italy, October 7-13, 2012, Proceedings, Part VI 12}.\hskip 1em plus 0.5em minus 0.4em\relax Springer, 2012, pp. 214--227.

\bibitem{rublee2011orb}
E.~Rublee, V.~Rabaud, K.~Konolige, and G.~Bradski, ``Orb: An efficient alternative to sift or surf,'' in \emph{2011 International conference on computer vision}.\hskip 1em plus 0.5em minus 0.4em\relax Ieee, 2011, pp. 2564--2571.

\bibitem{leutenegger2011brisk}
S.~Leutenegger, M.~Chli, and R.~Y. Siegwart, ``Brisk: Binary robust invariant scalable keypoints,'' in \emph{2011 International conference on computer vision}.\hskip 1em plus 0.5em minus 0.4em\relax Ieee, 2011, pp. 2548--2555.

\bibitem{ozuysal2009fast}
M.~Ozuysal, M.~Calonder, V.~Lepetit, and P.~Fua, ``Fast keypoint recognition using random ferns,'' \emph{IEEE transactions on pattern analysis and machine intelligence}, vol.~32, no.~3, pp. 448--461, 2009.

\bibitem{zhao2012flip}
W.-L. Zhao and C.-W. Ngo, ``Flip-invariant sift for copy and object detection,'' \emph{IEEE Transactions on Image Processing}, vol.~22, no.~3, pp. 980--991, 2012.

\bibitem{tareen2018comparative}
S.~A.~K. Tareen and Z.~Saleem, ``A comparative analysis of sift, surf, kaze, akaze, orb, and brisk,'' in \emph{2018 International conference on computing, mathematics and engineering technologies (iCoMET)}.\hskip 1em plus 0.5em minus 0.4em\relax IEEE, 2018, pp. 1--10.

\bibitem{norman2023actag}
K.~Norman, D.~Butterfield, and J.~G. Mangelson, ``Actag: Opti-acoustic fiducial markers for underwater localization and mapping,'' in \emph{2023 IEEE/RSJ International Conference on Intelligent Robots and Systems (IROS)}.\hskip 1em plus 0.5em minus 0.4em\relax IEEE, 2023, pp. 9955--9962.

\bibitem{tueller2018comparison}
P.~Tueller, R.~Kastner, and R.~Diamant, ``A comparison of feature detectors for underwater sonar imagery,'' in \emph{OCEANS 2018 MTS/IEEE Charleston}.\hskip 1em plus 0.5em minus 0.4em\relax IEEE, 2018, pp. 1--6.

\bibitem{lin2014microsoft}
T.-Y. Lin, M.~Maire, S.~Belongie, J.~Hays, P.~Perona, D.~Ramanan, P.~Doll{\'a}r, and C.~L. Zitnick, ``Microsoft coco: Common objects in context,'' in \emph{Computer Vision--ECCV 2014: 13th European Conference, Zurich, Switzerland, September 6-12, 2014, Proceedings, Part V 13}.\hskip 1em plus 0.5em minus 0.4em\relax Springer, 2014, pp. 740--755.

\bibitem{deng2009imagenet}
J.~Deng, W.~Dong, R.~Socher, L.-J. Li, K.~Li, and L.~Fei-Fei, ``Imagenet: A large-scale hierarchical image database,'' in \emph{2009 IEEE conference on computer vision and pattern recognition}.\hskip 1em plus 0.5em minus 0.4em\relax Ieee, 2009, pp. 248--255.

\bibitem{Soundmetrics}
\BIBentryALTinterwordspacing
Soundmetrics, \emph{Soundmetrics - Sonar providers}, Soundmetrics. [Online]. Available: \url{http://www.soundmetrics.com/Products}
\BIBentrySTDinterwordspacing

\bibitem{Teledynemarine}
\BIBentryALTinterwordspacing
teledynemarine, \emph{teledynemarine - Sonar providers}, teledynemarine. [Online]. Available: \url{https://www.teledynemarine.com/en-us/brands/Pages/blueview.aspx}
\BIBentrySTDinterwordspacing

\bibitem{Blueprintsubsea}
\BIBentryALTinterwordspacing
blueprintsubsea, \emph{blueprintsubsea - Sonar providers}, blueprintsubsea. [Online]. Available: \url{https://www.blueprintsubsea.com/oculus/}
\BIBentrySTDinterwordspacing

\bibitem{Tritech}
\BIBentryALTinterwordspacing
tritech, \emph{tritech - Sonar providers}, tritech. [Online]. Available: \url{https://www.tritech.co.uk/products/gemini-720is}
\BIBentrySTDinterwordspacing

\end{thebibliography}

\end{document}